\documentclass[twocolumn, secnumarabic,  floatfix, nofootinbib, nobalancelastpage, longbibliography]{revtex4-2}

\def\TitleOfPaper{Exchange-correlation energy from Green's functions}
\def\preprintlinklocation{http://dft.uci.edu}

\usepackage{amsmath}
\usepackage{physics}
\usepackage{amssymb}
\usepackage{float}
\usepackage[dvipsnames]{xcolor}
\usepackage{subcaption}
\usepackage{multirow}
\usepackage{caption}

\definecolor{TITLECOL}{rgb}{0.05,0.25,0.85}
\definecolor{CONTENTSCOL}{rgb}{0.1,0.2,0.7}
\definecolor{URLCOL}{rgb}{0,0.52,0.83}
\definecolor{LINKCOL}{rgb}{0.05,0.5,0}
\definecolor{CITECOL}{rgb}{0.25,0,0.48}
\definecolor{SECOL}{rgb}{0.07,0.31,0.80}
\definecolor{SSECOL}{rgb}{0.26,0.19,0.75}

\newcommand{\coloredtitle}[1]{\title{\textcolor{TITLECOL}{#1}}}
\newcommand{\coloredauthor}[1]{\author{\textcolor{CITECOL}{#1}}} 

\usepackage{graphicx}

\def\PublicationsLink{http://dft.uci.edu/publications.php}
\def\preprintlink{ \href{\preprintlinklocation}{\TitleOfPaper} }
\def\preprinttext{~}

\usepackage{fancyhdr}
\makeatletter
\fancypagestyle{titlepage}
{	\lhead{\textsc{\preprinttext\ ~ \href{\PublicationsLink}{~}}}
	\chead{}
	\rhead{\preprintlink}
	\lfoot{}}
\makeatother
\def\preprintlink{ 
	\href{\preprintlinklocation}
        {
~}
	}

\definecolor{Green}{rgb}{0.016,0.627,0}
\definecolor{Plum}{rgb}{0.17,0,0.45}
\definecolor{LBlue}{rgb}{0,0.34,0.45}
\definecolor{Sepia}{rgb}{0.37,0.17,0.02}
\definecolor{BurntOrange}{rgb}{0.78,0.39,0}

\usepackage{hyperref}
\hypersetup{
    breaklinks=true,
    bookmarksopen=true,
    bookmarksnumbered=true,
    colorlinks=true,
    linkcolor=LINKCOL,
    linktocpage=true,
    citecolor=CITECOL,
    filecolor=magenta,      
    urlcolor=URLCOL,
}

\def\bea{\begin{eqnarray}}
\def\eea{\end{eqnarray}}
\def\ben{\begin{equation}}
\def\een{\end{equation}}
\def\benu{\begin{enumerate}}
\def\enu{\end{enumerate}}

\def\bei{\begin{itemize}}
\def\eei{\end{itemize}}
\def\beit{\begin{itemize}}
\def\eit{\end{itemize}}
\def\benu{\begin{enumerate}}
\def\enu{\end{enumerate}}

\def\n{n}

\def\sss{\scriptscriptstyle\rm}

\def\1var{(\bx_1...\bx\N)}

\def\br{{\bf r}}

\def\bx{{\bf x}}

\def\x{_{\sss X}}
\def\c{_{\sss C}}
\def\s{_{\sss S}}

\def\xc{_{\sss XC}}

\def\Hxc{_{\sss HXC}}

\def\N{_{\sss N}}

\def\LDA{^{\rm LDA}}

\def\ee{_{\rm ee}}

\def\sph_int{ {\int d^3 r}}

\def\intr{\int d^3r\,}

\def\om{\omega}
\def\Msum{-i\int\frac{d \omega}{2\pi}}

\def\G{G}
\def\t{t}
\def\D{G}

\def\Exc{E\xc }

\DeclareMathOperator{\sign}{sgn}

\def\UEG{^{\rm UEG}}

\def\QP{^{\rm SP}}
\def\SAT{^{\rm MP}}
\def\QPI{^{\rm SPI}}
\def\STI{^{\rm MPI}}

\newcounter{edit}

\newcommand{\cmmnt}[1]{}
\graphicspath{{figures/}}

\begin{document}
\sf
\coloredtitle{\TitleOfPaper}

\coloredauthor{Steven Crisostomo}
\email{crisosts@uci.edu}
\affiliation{Department of Physics and Astronomy, University of California, Irvine, CA 92697, USA}

\coloredauthor{E.K.U. Gross}
\affiliation{Fritz Haber Center for Molecular Dynamics, Institute of Chemistry, The Hebrew University of Jerusalem, Jerusalem 91904, Israel}

\coloredauthor{Kieron Burke}
\affiliation{Department of Chemistry, University of California, Irvine, CA 92697, USA}
\affiliation{Department of Physics and Astronomy, University of California, Irvine, CA 92697, USA}
\date{\today}

\begin{abstract}
DFT calculations yield useful ground-state energies and densities, while Green's function
techniques (such as $GW$) are mostly used to produce spectral functions.
From the Galitskii-Migdal formula, we extract the exchange-correlation of DFT
directly from a Green's function.  This
spectral representation  provides an alternative to the fluctuation-dissipation theorem of DFT, identifying distinct single-particle and many-particle contributions. Results are illustrated on the uniform electron gas and the two-site Hubbard model.

\end{abstract}
\maketitle

Ground-state density functional theory \cite{HK64,KS65} (DFT) is used to great effect in modern
molecular and materials calculations, limited
only by the approximation for the exchange-correlation (XC) energy \cite{K99}.
But there is also interest in the response properties of
a system, such as its spectral function and associated gap \cite{specfunc1}.   While range-separated hybrids
can yield moderately accurate gaps within the generalized KS scheme \cite{S96,P17,GKS20}, the standard method
for gaps remains
$GW$ calculations \cite{H65,GDR19}.  For strongly-correlated materials, dynamical mean field theory has become
a very useful tool \cite{DMFT1,DMFT2}.

Such Green's function (GF) methods come from traditional many-body theory.  KS-DFT does not, and it is difficult to relate the two. This leads to issues whenever, as is common practice, a DFT calculation is used to find
some set of KS orbitals, which form the starting point of GF calculations \cite{GW100_2015}.
The choice of initial orbitals can be important \cite{BM13,L18,WOTLeeor22},
while a self-consistent GF can eliminate this dependence \cite{vSKF06,QPGW14,SCQPGW21,PYIZG22}. The XC energy of DFT is traditionally analyzed in terms of static quantities extractable from the ground-state wave-function, such as the XC hole, the
dearth of conditional electronic density around an electron \cite{B88,PBE96,PBE099}.  If a frequency decomposition is performed, it is in terms of the density-density response function \cite{HJ74}.

We provide a bridge between these two distinct approaches by extracting the contribution to the XC energy directly from an interacting GF.
Our GXC formula can be used to (a) extract approximate XC energies from approximate
GF, (b) provide a novel decomposition of XC energies into single-particle and many-particle contributions, (c) provide models for existing DFT approximations in terms of GF, (d) relate
the quality of spectral functions to their performance for XC energies. 
We illustrate our results on the uniform 
electron gas and the two-site Hubbard model.
 
Consider Hamiltonians of the form (in Hartree atomic units)
\ben \hat{H} = \sum_{j}^{N}\hat{h}(\br_{i}) + \sum_{i>j}^{N}\hat{V}\ee(|\br_{i}-\br_{j}|), \label{eq:Ham} \een
where $\hat{h}(\br) = \hat{t}(\br) + v(\hat{\br})$ is the single-particle Hamiltonian, with kinetic energy operator $\hat{t} = -\nabla^{2}/2$, multiplicative spin-independent external potential $v(\hat{\br})$, and $\hat{V}\ee(u)=1/u$ is the Coulomb repulsion between pairs of electrons. 
We work within the Born-Oppenheimer approximation in the non-relativistic limit, with no external magnetic fields.
We use $\bx = (\br,\sigma)$ as a space-spin coordinate, and $n(\bx)$ is the ground-state spin density.

Within spin-DFT, the corresponding KS system consists of fictitious non-interacting electrons with the same ground-state density as the interacting system.
The exact ground-state energy is then 
\ben E = T\s + V + E\Hxc, \label{eq:EKS1}\een
where $T\s$ is the kinetic energy of non-interacting electrons, $V$ is the external potential energy, and $E\Hxc$ is the sum of the Hartree
(electrostatic) and XC energies.   The KS potential is
\ben v\s(\bx)=v(\br)+v\Hxc(\bx), \een
where
$v\Hxc(\bx) = \delta E\Hxc[n]/\delta n(\bx)$. The KS equations are solved self-consistently, yielding the exact density and energy given the exact XC \cite{SWWB12}.

The time-ordered Green's function is
\ben \G(\bx \t ,\bx' \t') = -i\big\langle \mathcal{T}\big[\hat{\psi}(\bx \t) \hat{\psi}^{\dag}(\bx' \t')\big] \big\rangle, \een
where $\mathcal{T}$ is the time-ordering operator and $\langle\cdot \rangle$ denotes the expectation value over the $N$-electron ground-state $\vert \Psi_{0}\rangle$. The fermionic operators, $\hat{\psi}^{\dag}(\bx)$ and $\hat{\psi}(\bx)$, create and destroy a particle of spin $\sigma$ at $\br$ and are time-evolved in the Heisenberg picture according to $ \hat{H}$. 
For time-independent $\hat H$, we denote the Fourier transform $G(\bx,\bx',\om)$. 
We define
\ben \Tr \{F\} = \Msum \tr\{F\} , \een
where a convergence factor $\exp(i\omega \delta)$ with $\delta \to 0^{+}$ is implied, to respect time-ordering ($\t'\to \t^{+}$). Here $\tr$ denotes
\ben \tr\{F\} = \intr \sum_{\sigma = \sigma'} \lim_{\br' \to \br} F(\bx,\bx',\om). \een
The Galitskii-Migdal (GM) formula \cite{GM58} yields the total interacting ground-state energy from
the GF
\ben E = \frac{1}{2}\Tr\{ (\om + \hat{h}(\br))\G \}.   \label{eq:GM1}\een

To isolate XC, apply GM to the KS system,
\ben E\s = \frac{1}{2}\Tr \{(\om  + \hat{h}\s(\bx))\G\s \} .\label{eq:GM2} \een
Subtraction yields GXC, the GF XC contribution
\ben \D\xc= \Exc  - \frac{ \langle v\xc \rangle }{2}  = \frac{1}{2}\Tr\left\{ \left(\om + \hat{t}(\br)\right)\Delta \G \right\} , \label{eq:GXC1}  \een
where $\Delta \G = \G - \G\s$.  More explicitly
\ben G\xc = -\frac{i}{2}\int d^{3}r \lim_{u\to 0} \int \frac{d\omega}{2\pi} \left(\omega -\frac{\nabla^{2}_{\textbf{u}}}{2}\right)\Delta G(\br,\br+\textbf{u},\omega), \label{eq:GXCh}\een
where $\Delta G(\br,\br+\textbf{u},\omega)$ is an analog of the XC hole. Like the XC hole, obeying sum rules. Without the $\omega$ term, Eq.~(\ref{eq:GXCh}) yields $T\c$.  Moreover
\ben -i\int \frac{d\omega}{2\pi}\,\Delta G(\br,\br,\omega) = 0,  \label{eq:SSE1}\een
because both Green's functions have the same density. 
An early version appeared \cite{S85} even before the adiabatic connection formula of DFT was precisely defined \cite{HJ74,LP75}. 

First, Eq.~(\ref{eq:GXC1}) provides a method for extracting an XC contribution directly from any GF.  At the end of any GF calculation, the density
of $G$ can be extracted, a KS inversion \cite{NS21,NM21,SW21,SCW22,VXC_solids_2023} performed and the corresponding $G\s$ constructed.   But $\D\xc$ is not $E\xc$.   Given an explicit approximation for $E\xc$, it is easy to construct $\D\xc$, but not vice versa.   Thus, for a non-self-consistent
$GW$ calculation (the vast majority), a measure of inconsistency would be $\D\xc$ of the original DFT calculation versus that of Eq.~(\ref{eq:GXC1}).
A self-consistent calculation would presumably satisfy Eq.~(\ref{eq:GXC1}) on its final iteration.
An important feature of $\D\xc$ is that it can be used to construct the total energy from the sum of KS orbital energies $E\s$:
\ben E = E\s + \left(\D\xc -\frac{\langle v\Hxc \rangle }{2}\right) .\een
Thus $\D\xc$ is related to a double-counting correction. 

Now we use Eq.~(\ref{eq:GXC1}) to create a novel decomposition of XC energies in DFT.
Define the energy difference
\ben \omega_{J} = E_{0}(N) - E_{J}(N-1),\een
where $E_{J}(M)$ is the energy of the $J$th interacting eigenstate of the $M$-electron system; $J=0$ denotes the ground-state. Then the Lehmann representation is
\ben G(\bx,\bx',\om) = \sum_{J}\frac{\rho_{J}(\bx,\bx')}{\om -\omega_{J} -i\delta} + \sum_{J'}\frac{\bar{\rho}_{J'}(\bx,\bx')}{\om -\bar{\omega}_{J'} +i\delta}, \label{eq:Lehmann1}\een
where the spectral weights are
\ben \rho_{J}(\bx,\bx') = \langle \Psi_{0} \vert \hat{\psi}^{\dag}(\bx') \vert J \rangle \langle J \vert \hat{\psi}(\bx) \vert \Psi_{0} \rangle. \label{eq:specweight} \een
Here $\{\vert J \rangle\}$ are the interacting eigenstates of the $N\text{-}1$-electron system, while 
$\bar{\rho}_{J'}(\bx,\bx')$ is defined analogously, with fermionic operators in Eq.~(\ref{eq:specweight}) swapped, and $\{\vert J'\rangle\}$ enumerating eigenstates of the $N+1$-electron system. The sum over $J$ of $\rho_{J}(\bx,\bx')$ is simply the first-order density matrix.
 Likewise, we write
\ben  G\s(\bx,\bx',\om) = \sum_{j}\frac{\rho_{s,j}(\bx,\bx')}{\om - \epsilon_{j} - i\delta} +  \sum_{j'}\frac{\bar{\rho}_{s,j'}(\bx,\bx')}{\om - \bar{\epsilon}_{j'} + i\delta}, \een
where $j$ runs over KS orbitals and we consider only the occupied orbitals. 
Inserting into Eq.~(\ref{eq:GXC1}) 
yields
\ben \D\xc = \frac{1}{2}\sum_J (T_J +\omega_{J}f_{J} ) - \sum_j (T_{s,j}+ \epsilon_{j}f_{s,j} ), \label{eq:GXC2} \een
where $T_J$ is the kinetic contribution from state $J$ and $f_J= \tr\{\rho_J\}$ are the 'occupations' of each state,
with analogous contributions from the KS system.  They satisfy the sum rule that each, when summed over all occupied states,
yields $N$.

To make further progress, we relate the two sums in Eq.~(\ref{eq:GXC2}) via the adiabatic connection of DFT.  Multiply $V\ee$ in Eq.~(\ref{eq:Ham}) by $\lambda$ while choosing a $\lambda$-dependent one-body potential to keep the density fixed \cite{SCP03}.  As $\lambda\to 0$, a subset of $J$ approaches the KS orbitals and eigenvalues.  We call these single-particle contributions (SP), the rest many-particle (MP). We denote sums over such excitations by $K$.  This is analogous to how we define single-particle excitations in TDDFT \cite{M22}.
Thus
\ben \D\xc = \D\xc\QP + \D\xc\SAT, \label{eq:GXC3} \een 
where the SP contribution is
\ben \D\QP\xc = \frac{1}{2}\sum_j (T_{{\sss C},j} +\omega_{j}f_{j} - \epsilon_{j}f_{{\rm s},j} ), \label{eq:GXCQP1} \een
and $T_{{\sss C},j} = T_{j}-T_{{\rm s},j} = \tr\{t(\br)(\rho_{j}-\rho_{{\rm s},j})\}$ is the contribution to the correlation kinetic energy from each SP state.  
The many-particle contributions are purely correlation
\ben \D\SAT\c = \frac{1}{2}\sum_K (T_{K} +\omega_{K} f_{K}).\label{eq:GXCSAT1} \een

In the SP contribution, the highest occupied KS level is special, as
KS-DFT guarantees that the HOMO KS eigenvalue is exactly minus the ionization potential, i.e.,
$\omega_{0} = -I = \epsilon_{0}$.  Labelling it as zero, we write
\ben \D\QP_{{\sss XC},0} =  \frac{1}{2}(T_{{\sss C},0} -I(f_{0}-f_{{\rm s},0}) ). \een

The correlation kinetic energy, $T\c$, is just the sum of its SP and MP contributions. 
We can analogously isolate contributions to the potential correlation energy.  Their sum yields the correlation energy, so
\ben E\xc = E\xc\QP + E\c\SAT, \een
i.e. individual peaks in the spectral function yield individual contributions to the $E\xc$ energy.
The exchange contribution is found by using the exchange self-energy (see SM), yielding
\ben \D_{\sss X} = \sum_j \tr \{ (\Sigma\x[G\s]-v\x)\rho_{{\rm s},j} \} = \D_{\sss X}\QP.\een
For a two-electron singlet $\D_{\sss X}$ vanishes.

Likewise, the ground-state density matrix is
\ben \rho(\bx,\bx') =\int^{0}_{-\infty} d\omega \,  A(\bx,\bx',\omega), \label{eq:specf}\een
where the spectral function is
\ben A(\bx,\bx',\omega) = -\frac{1}{\pi}{\rm Im}\{G(\bx,\bx',\omega)\}\sign(\omega). \een
Thus $\rho$ (and its diagonal, the ground-state density) also can be uniquely decomposed.  


An alternative decomposition uses
$\omega_{j} = -I - \Delta E_{j},$
where $\Delta E_{j} = E_{j}(N-1)-E_{0}(N-1) \geq 0$ are the transition frequencies.  Then
\ben \D\xc = \D\xc\QPI + \D\c\STI,\label{eq:GXC4} \een
with different SP and MP contributions
\ben \D\QPI\xc = \frac{1}{2}\sum_{j}\left[T_{c,j} - (\Delta E_{j}f_{j}-\Delta E_{s,j}f_{s,j}) \right], \label{eq:GXCQP2}\een
\ben \D\STI\c =  \frac{1}{2}\sum_{K} \left[T_{K} -\Delta E_{K} f_{K}\right].\label{eq:GXCSAT2}\een
In particular, $\D\QPI_{xc,0} = T_{c,0}/2$.  
Each decomposition is useful in different circumstances.

The asymmetric two-site Hubbard model \cite{Carrascal_2015,BK21} is particularly well-suited as an illustration, because of its extremely truncated Hilbert space.  We have two femions in the Hamiltonian,
\ben -t\sum_{\sigma}\big(\hat{c}^{\dag}_{1\sigma}\hat{c}_{2\sigma} + \text{h.c.}\big) + U\sum_{j=1}^{2}\hat{n}_{j\uparrow}\hat{n}_{j\downarrow} + \sum_{j=1}^{2}v_{j}\hat{n}_{j}, \een
where $t$ is the hopping term, $U$ the on-site interaction strength, $\hat{n}_{j}$ the site-occupation operator, and $\hat{c}_{j\sigma}^{\dag}$, $\hat{c}_{j\sigma}$ are the fermionic operators associated to each site. Only
the potential difference $\Delta v = v_{2}-v_{1}$ matters.  The ground-state is a singlet, and its density
is characterized by one number, $\Delta n = n_{2} - n_{1}$. KS-DFT applies \cite{Carrascal_2015} and the KS system is simply the tight-binding dimer.

\begin{figure}[H]
\includegraphics[width=\linewidth]{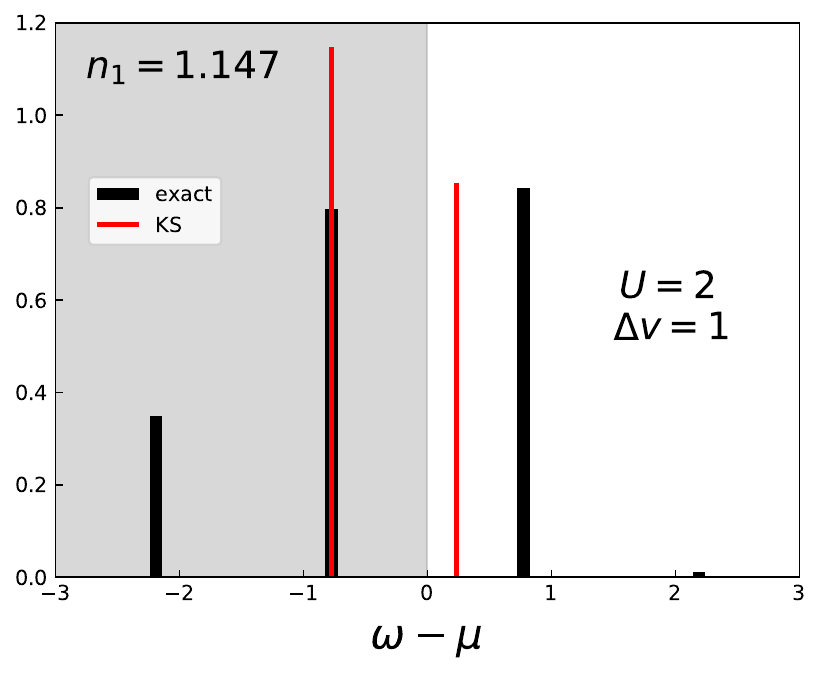}
\caption{Hubbard dimer local spectral functions $A_{11}(\omega)$ (black) and $A_{{\rm s},11}(\omega)$ (red). Here $\mu = U/2$ and $t=1/2$. The spectral weight below 0 sums to $n_{1}= 1.147$, and above to $n_2$.}
\label{fig:1}
\end{figure}

Figure \ref{fig:1} shows a typical spectral function \cite{Carrascal_2015} with bars whose height is the pole weight. $G$ has 4 poles, two for removal and two for gain, but $G\s$ has only one of each.  The difference between the two central peaks is the gap, with the KS gap famously being smaller than the true gap \cite{PL83,SS83}.
Only the removal peaks contribute to XC.  The larger exact peak is the only single-particle contribution, and the smaller is a many-particle.  By Eq.~(\ref{eq:specf}), they sum to $n_1$, and the KS peak must be higher.  The many-particle contributes negatively to $\D\xc$ while
SP contributes positively, a delicate balance.

\begin{table}[h]
  \centering
  \begin{tabular}{@{}|c|c|c|c|c|c|c|c|c|c|@{}}
    \hline
    $U$ & $ E\xc$ & $E\c$ & $\D\xc$ & $\D\xc\QP$ & $ \D\xc\SAT $ & $ \D\xc\QPI$ & $ \D\xc\STI$    \\
    \hline
   0.5 & -0.339 & -0.01062 & -0.013 & 0.011 & -0.024 & 0.00338 & -0.0164 \\
   \hline
   1 & -0.643 & -0.0676 & -0.0524 & 0.0516 & -0.104 & 0.0189 & -0.0713 \\
   \hline
   2 & -1.39 & -0.3666 & -0.139 & 0.224 & -0.363 & 0.0847 & -0.224 \\
   \hline
   4 & -3.23 & -1.224 & -0.194 & 0.689 & -0.883 & 0.188 & -0.381 \\
   \hline
   10 & -9.1 & -4.098 & -0.206 & 2.16 & -2.36 & 0.27 & -0.476 \\
   \hline 
   20 & -19. & -9.05 & -0.207 & 4.64 & -4.85 & 0.297 & -0.504 \\
   \hline
  \end{tabular}
  \caption{XC of dimer with $t=1/2$ and $\Delta v = 1$. }
  \label{tab:GXCexactV1}
\end{table}

The first use of our formula is to identify how much each peak contributes to XC. Table \ref{tab:GXCexactV1}  shows values of various quantities for a range of $U$. $\G\xc$ is comparable to $E\c$ for weak correlation (as $\G\x=0$ for $N=2$), but significantly smaller for strong correlation, with cancellation between SP and MP contributions. The alternative decomposition yields terms of much smaller magnitude, requiring a less delicate balance, especially for large $U$.


\begin{table}[h]
  \centering
  \begin{tabular}{|c|c|c|c|c|c|}
    \hline
    $Z$ & $E\xc $ & $ \langle v\xc\rangle/2 $ &  $ E\c$ & $T\c$  &  $ \D\xc$ \\
    \hline
    1 & -0.4229 & -0.3562 & -0.04199 & 0.02788 & -0.06667 \\
    \hline
    2 & -1.067 & -1.01 & -0.04211 & 0.03664 & -0.05691 \\
    \hline
    3 & -1.695 & -1.638 & -0.04352 & 0.03983 & -0.05627 \\
    \hline
    4 & -2.321 & -2.265 & -0.04427 & 0.04148 & -0.05604 \\
    \hline
    6 & -3.572 & -3.516 & -0.04506 & 0.04318 & -0.05584 \\
    \hline
    10 & -6.073 & -6.017 & -0.04569 & 0.04456 & -0.0557 \\
    \hline
    20 & -12.32 & -12.27 & -0.04618 & 0.0456 & -0.0556 \\
    \hline
  \end{tabular}
  \caption{XC for the two-electron ions \cite{HU97}. }
  \label{tab:2eion}
\end{table}

For realistic DFT calculations, as the local density approximation yields $E\x\LDA[n] = -A\x\int n^{4/3}(\br)$, typically $\D\xc \approx E\xc/3$, especially for large $N$.  For $N=2$, we cannot assume Hubbard results are typical, but (essentially) exact DFT calculations have been performed for two-electron ions (Table \ref{tab:2eion}) and Hooke's atom (quadratic confining potential in the SM), where $\G\c$ is comparable to $E\c$ in magnitude. 

The LDA can be understood as a local approximation to the XC hole \cite{GL76,GJ80}, as the system- and spherically averaged LDA XC hole reflects the accuracy of LDA \cite{GJL77}. The real-space construction of the GGA from the gradient expansion for the hole underlies both the PW91 and PBE XC approximations \cite{BPW98,PBE96}. XC holes
are also behind some of the most popular functionals in chemistry \cite{B88,BPW98,PBE099}.  Now we derive $\D\xc\LDA$ from an ansatz for the Green's function.   Define $\Delta G\UEG(n,u,\om)$ as the difference between exact and KS Green's functions of
a uniform gas of density $\n$, separation $u = \vert \br - \br' \vert$, and frequency $\om$.  Approximating $\Delta G$ with $\Delta G\UEG(\n(\br),|\br-\br'|,\omega)$ in Eq.~(\ref{eq:GXC1}) directly yields
\ben 
\D\xc\LDA = \int d^{3}r\, g\xc\UEG(n(\br)), 
\label{eq:GXCLDA}
\een
where $g\xc\UEG(n)= (2- d/dn)e\xc\UEG(n)/2$ is the $\D\xc$ energy density, $\Tr\{(\omega + \hat{t}(u))\Delta G\UEG\}/V$, and $V$ is the volume. 
System-averaged and frequency-integrated quantities should agree to the extent that LDA yields reasonably accurate energies, but are there major cancellations of errors inside the frequency integral?   And do approximate GF calculations improve this frequency dependence? These are the sorts of questions that can be explored with our formula.

Our final point concerns the approximate GF calculations that our formula is designed to analyze,  illustrated for $GW$ calculations on the Hubbard dimer.  
We use the Hartree-Fock GF to generate the initial $GW$ self-energy and iterate until the energy convergences. Each iteration generates extra poles, but we retain only a few (see SM).  If the self-energy has (correctly) only two poles, the next $G^{GW}$ generally has six, instead of the correct four. But if we reduce the 6 poles of $G$ to the correct 4, the self-energy has 4.  We chose the former scheme for calculations here. To aid convergence, for cases with moderate to strong correlation we generated the $n$th GF according to: $G^{(n)} = \gamma G^{(n-1)} + (1-\gamma) G^{(n-2)}$, with $\gamma = 0.67$. Our one-shot results are identical to Ref \cite{RGR09} when $\Delta v = 0$.

\begin{figure}[h]
\includegraphics[width=\linewidth]{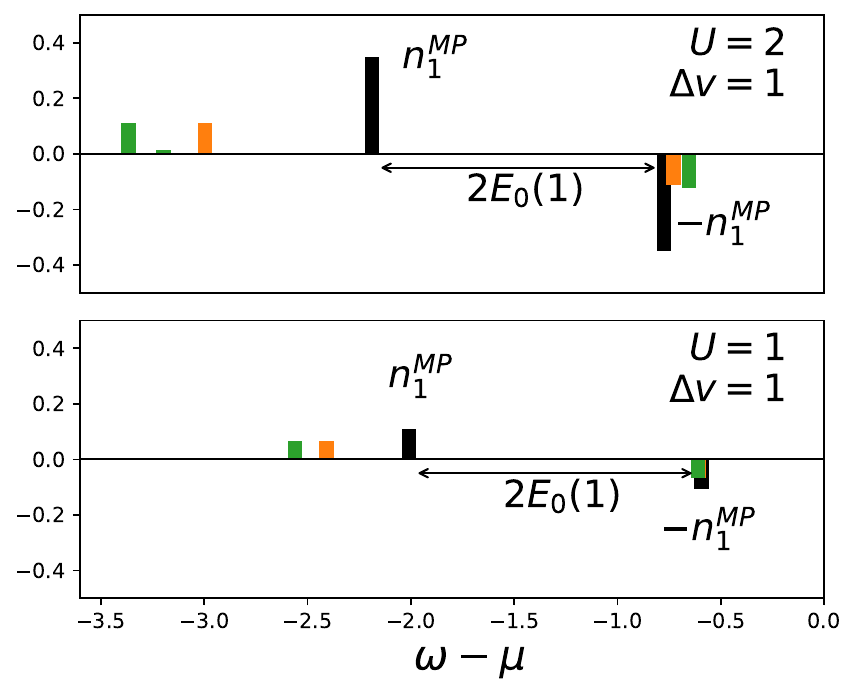}
\caption{Spectral function differences $\Delta A_{11}(\omega)$ with $\mu = U/2$ and $t=1/2$. Exact (black), one-shot (orange), and self-consistent $GW$ (green). The separation of the SP and MP poles is twice the energy of the one-particle groundstate, but $GW$ introduces an erroneous interaction dependence.  }
\label{fig:2}
\end{figure}

Both the six-pole and four-pole $G^{GW}$ yield the same density and $v\xc$, but they differ considerably. Both satisfy the Sham-Schl{\"u}ter equation at each iteration \cite{SS83},
\ben -i\int \frac{d\omega}{2\pi}\big[ G\s\Sigma G\big]_{ii} = -i\int \frac{d\omega}{2\pi}\big[ G\s v\Hxc G \big]_{ii},\een
but have different kinetic and XC energies. For the dimer
\ben \D\xc = \int_{-\infty}^{0} d\omega \left[\omega (\Delta A_{11}(\omega)+\Delta A_{22}(\omega)) - 2t \Delta A_{12}(\omega)\right], \een
where we have summed over spins and $\Delta A_{ij}$ is the difference in retarded spectral functions.
Due to Eq.~(\ref{eq:SSE1}), the frequency integral of $\Delta A_{11}(\omega)+\Delta A_{22}(\omega)$ up to the chemical potentials yields 0. Fig.~\ref{fig:2} shows $\Delta A_{11}$, which comprises a majority of $G\xc$, as $\Delta A_{22}$ gives a negligible contribution.

From Fig.~\ref{fig:2}, the SP pole positions and heights for one-shot $GW$ are more accurate than the self-consistent $GW$. Table~\ref{tab:tab4} also shows a poorer approximation to $E\c$, worsening with increasing $U$. However, full self-consistency produces better total energies for weak correlation, while in the strongly-correlated case, neither flavor of $GW$ produces accurate energies (see SM). 

\begin{table}[h]
  \centering
  \begin{tabular}{|c|c|c|c|c|c|c|c|c|c|c|c|c|c|c|c|}
    \hline
    \multicolumn{1}{|c}{} &
    \multicolumn{3}{c|}{Exact} &
    \multicolumn{3}{|c|}{one-shot} &
    \multicolumn{3}{|c|}{self-consistent} \\   
    \hline
    $U$  &  $\D\xc$ & $ \D\xc\QPI $ & $ \D\xc\STI$ &  $\D\xc$ & $ \D\xc\QPI $ & $ \D\xc\STI$ &  $\D\xc$ & $ \D\xc\QPI $ & $ \D\xc\STI$   \\
    \hline
   0.25 &  -3.05 & 0.67 & -3.72 &  -4.88 & 1.12 & -6.  & -4.95 & 1.15 & -6.1 \\
   \hline
   0.5 &  -13. & 3.38 & -16.4 &   -15.9 & 4.4 & -20.3  & -16.9 & 4.69 & -21.6 \\
   \hline
   1 &  -52.4 & 18.9 & -71.3 &   -41.9 & 14.4 & -56.3  & -54.8 & 16.2 & -71.  \\
   \hline
   2 &  -139. & 84.7 & -224. &  -98.1 & 33.8 & -132. & -186. & 42. & -228. \\
   \hline
   4 & -194.\phantom{0} & 188. & -381. &  -232. & 57.5 & -289. & -583. & 88. & -671. \\
   \hline
  \end{tabular}
  
\vspace*{0.25 cm}

  \begin{tabular}{|c|c|c|c|c|c|c|c|c|c|c|c|c|c|c|c|}
    \hline
    \multicolumn{1}{|c}{} &
    \multicolumn{3}{c|}{Exact} &
    \multicolumn{3}{|c|}{one-shot} &
    \multicolumn{3}{|c|}{self-consistent} \\   
    \hline
    $U$  &  $E\c$ & $ E\c\QPI $ & $ E\c\STI$ &  $E\c$ & $ E\c\QPI $ & $ E\c\STI$ &  $E\c$ & $ E\c\QPI $ & $ E\c\STI$   \\
    \hline
   0.25 & -1.95 & 1.76 & -3.71 &  -3.09 & 2.88 & -5.97  & -3.83 & 2.25 & -6.08 \\
   \hline
   0.5 &  -10.6 & 5.62 & -16.2 &   -13.8 & 6.29 & -20.1  & -19.3 & 2.18 & -21.5 \\
   \hline
   1 &  -67.6 & 1.43 & -69. &    -68.1 & -13.3 & -54.9  & -87.9 & -16.5 & -71.4  \\
   \hline
   2 &  -367. & -142. & -225. &  -288. & -153. & -135. & -292. & -56.3 & -236. \\
   \hline
   4 & -1220. & -627. & -597. &  -798. & -463. & -335. & -761. & -60. & -701. \\
   \hline
   
  \end{tabular}
\caption{Decompositions for the asymmetric dimer with $t=1/2$ and $\Delta v = 1$, in milliHartrees.}
\label{tab:tab4}
\end{table}

Practical GF calculations are performed on solids and molecules, with a variety of approximations, such as $GW$ and dynamical mean field theory \cite{F92,FHK96,GLSRGKRSSR11,KLAKKV17,ZKP20}.
Once a KS inversion can be performed \cite{SW21,SCW22,VXC_solids_2023} on the density of the approximate GF, a value for $G\xc$ can be extracted which can be compared with a DFT counterpart, especially when standard XC functionals are known to fail.  However, molecules have both bound and continuum states, while all states are in continuua in solids,  complicating the identification of SP and MP contributions.  For $GW$, there are many different recipes yielding distinct spectral features \cite{Interacting_Electrons_2016,vSKF06,ADvL09,L18}, but one must always use the KS GF of the density of the approximate GF, as the XC contributions depend on delicate cancellations, as illustrated here.
 Recently Ref.~\cite{ESR24} shows that exact XC energies can be produced via approximate self-energies. Our work is complementary, with our focus primarily being representations of XC energies via approximate GF. Our systematic determination of sources of XC errors should aid in designing future alternative approximate density functionals or GF methods.

We thank Lucia Reining, Abdallah El-Sahili, and Vojt\v{e}ch Vl\v{c}ek for valuable discussions. S.C. and K.B. were supported by NSF Award No. CHE-2154371. E.K.U.G acknowledges support form the European Research Council. This project has received funding from the European Research Council (ERC) under the European Unions Horizon 2020 research and innovation programme (Grant Agreement No. ERC-2017-AdG-788890).

\bibliographystyle{apsrev4-2}
\bibliography{main}

\label{page:end}
\end{document}


\sf
\coloredtitle{\TitleOfPaper}

\coloredauthor{Steven Crisostomo}
\email{crisosts@uci.edu}
\affiliation{Department of Physics and Astronomy, University of California, Irvine, CA 92697, USA}

\coloredauthor{E.K.U. Gross}
\affiliation{Fritz Haber Center for Molecular Dynamics, Institute of Chemistry, The Hebrew University of Jerusalem, Jerusalem 91904, Israel}

\coloredauthor{Kieron Burke}
\affiliation{Department of Chemistry, University of California, Irvine, CA 92697, USA}
\affiliation{Department of Physics and Astronomy, University of California, Irvine, CA 92697, USA}
\date{\today}

\maketitle

Below we detail various results for the two-site Hubbard model (Hubbard dimer). Our results for the dimer include: the exact tight-binding and interacting dimer Green's functions, one-shot $GW$ self-energy and Green's function beginning from Hartree-Fock. We report exact and approximate $G\xc$ (with SP and MP decompositions) for several dimer parameters. Additionally, we detail our self-consistent $GW$ method, and show how it compares with results reported in the literature. Lastly, we report $G\xc$ for the Coulombically interacting Hooke's atom system. 

\sec{Two-site tight-binding GF}
\label{sec:TBGF}
For any two-site tight-binding model,
\ben \hat{H}\s = -t\sum_{\sigma}\big(\hat{c}^{\dag}_{1\sigma}\hat{c}_{2\sigma} + \text{h.c.}\big) + \sum_{j=1}^{2} v_{{\rm{s}}, j}\hat{n}_{j}, \een
with hopping $t$ and on-site potential $v_{{\rm{s}},j}$, the ground-state time-ordered GF has an expression in terms of its site-occupations
\ben G\s(\omega) = \frac{\rho\s(n_{1},n_{2})}{\omega -\epsilon\HOMO -i\delta} + \frac{\bar{\rho}\s(n_{1},n_{2})}{\omega -\epsilon\LUMO +i\delta}. \een
Here the HOMO/LUMO eigenvalues $\epsilon_{\pm}$ are exactly
\ben \epsilon_{\pm} = (\bar{v}\s - \mu\s) \mp t\s, \een
where $\mu\s$ is the KS chemical potential. We define the average on-sight potential $\bar{v}\s = (v_{\rm s,2} + v_{\rm s,1})/2$, the rescaled hopping parameter $t\s = \sqrt{t^{2} + (\Delta v\s/2)^{2}}$, and the on-site potential difference $\Delta v\s = v_{{\rm s},2} - v_{{\rm s},1}$.
The amplitude matrices are function(al)s of the density, the first being the KS one-body reduced density (RDM)
\ben
	\rho\s(n_{1},n_{2}) =
	\frac{1}{2}\left( {\begin{array}{cc}
	n_{1} & \sqrt{n_{1}n_{2}}  \\
  \sqrt{n_{1}n_{2}} & n_{2}   \\
   \end{array} } \right),
\een
and $\bar{\rho}\s(n_{1},n_{2}) = \sigma_{y}\rho\s(n_{1},n_{2})\sigma_{y}$, using the Pauli $\sigma_{y}$ matrix. 

The KS kinetic energy is
\ben T\s(n_{1},n_{2}) = -2t\sqrt{n_{1}n_{2}}, \een
and for $N=2$ groundstate,
\ben -2\frac{d T\s(n_{1},n_{2})}{ d (\Delta n)} = \Delta v\s = \frac{-2t\Delta  n}{\sqrt{4-\Delta n^{2}}}. \een
The ionization potential theorem (IPT) requires $\epsilon\HOMO = -I $, minus the ionization potential, so
\ben \bar{v}\s  = \bar{v} -I + t\s,\een
where we have absorbed the exact and KS chemical potentials into $\bar{v}$ and $\bar{v}\s$, respectively. For both exact and approximate KS potentials, we fix the constant such that the IPT is satisfied.

\sec{Exact GF of Hubbard dimer}

The groundstate for the dimer can be found by directly diagonalizing the Hamiltonian in the two-particle singlet subspace. Defining $\bar{v} = (v_{1}+v_{2})/2 $:
\ben
	H =
	\left( {\begin{array}{cccc}
	U + 2\bar{v}-\Delta v & -t & -t & 0 \\
   -t & 2\bar{v} & 0 & -t  \\
   -t & 0 & 2\bar{v} & -t  \\
    0 & -t & -t &  U +2\bar{v}+ \Delta v \\
   \end{array} } \right),
\een
where the two-electron groundstate is non-magnetic and has the generic form \cite{Carrascal_2015}
\begin{multline} \ket{\Psi_{0}} = \alpha(t,U,\Delta v) \ket{1\uparrow 1\downarrow} + \bar{\alpha}(t,U,\Delta v) \ket{2\uparrow 2\downarrow} \\+ \beta(t,U,\Delta v)(\ket{1\uparrow 2\downarrow} + \ket{2\uparrow 1\downarrow}), \end{multline}
where the amplitudes depend on the system parameters and $\bar{\alpha}(t,U,\Delta v) = \alpha(t,U,-\Delta v) $ \cite{Carrascal_2015}.
    
In the Lehmann representation,
\ben \rho\QS_{ij} = \langle \Psi_{0} \vert \hat{c}_{j}^{\dag} \vert \psi^{(1)}_{0,1} \rangle \langle\psi^{(1)}_{0,1} \vert \hat{c}_{i} \vert \Psi_{0}\rangle, \label{eq:QS} \een 
where $\ket{\psi^{(1)}_{0,1}}$ are the ground-state and only excited state of the one-electron space. The spectral weights are thus,
\ben \rho\QS = M_{0} \mp (M_{1}- M_{2}), \een
with matrics defined via the wave-function coefficients:
\ben
	M_{0} =
	\frac{1}{2}\left( {\begin{array}{cc}
	\alpha^{2} + \beta^{2} & \beta(\alpha + \bar{\alpha})  \\
   \beta(\alpha + \bar{\alpha}) & \bar{\alpha}^{2} + \beta^{2}   \\
   \end{array} } \right),
\een

\ben
	M_{1} =
	\frac{t}{2\bar{t}}\left( {\begin{array}{cc}
	(\beta^{2} -\alpha^{2} )\frac{\Delta v}{2t}  & \beta(\bar{\alpha} - \alpha )\frac{\Delta v}{2t}   \\
   \beta(\bar{\alpha} - \alpha )\frac{\Delta v}{2t} & (\bar{\alpha}^{2} - \beta^{2}  )\frac{\Delta v}{2t}   \\
   \end{array} } \right),
\een

\ben
	M_{2} =
	\frac{t}{2\bar{t}}\left( {\begin{array}{cc}
	2\alpha \beta & (\beta^{2}+\alpha \bar{\alpha})  \\
   (\beta^{2}+\alpha \bar{\alpha}) & 2\bar{\alpha} \beta   \\
   \end{array} } \right),
\een
where the rescaled hopping is $\bar{t} = \sqrt{t^{2} + (\Delta v/2)^{2}}$. 

The spectral weights sum to the interacting density matrix,
\ben \rho = \rho\QP + \rho\SAT, \een
and we define the separate contributions to the density,
\ben n\QS_{i} = \rho\QS_{ii},  \een
where the subscripts $0,1$ in Eq.~(\ref{eq:QS}) denote SP or MP and $i$ denotes the site. Using the non-interacting functional $\rho\s(n_{1},n_{2})$ we express the $\rho\QS$ matrices as functionals of the single-particle and many-particle occupations,
\ben \rho\QP = \rho\s(n\QP_{1},n\QP_{2}), \quad \quad \quad \rho\SAT = \bar{\rho}\s(n\SAT_{2},n\SAT_{1}), \een
and by extension:
\begin{multline} G(\omega) = \frac{\rho\QP}{\omega - \omega\QP - i\delta} + \frac{\rho\SAT}{\omega - \omega\SAT - i\delta} \\ + \frac{\bar{\rho}\QP}{\omega + \omega\QP + i\delta} + \frac{\bar{\rho}\SAT}{\omega + \omega\SAT + i\delta}, \end{multline}
with SP and MP pole locations
\ben \omega\QS = E_{0}^{(2)} \pm \bar{t} -\frac{U}{2}, \een
where $E_{0}^{(2)}$ is the energy of the two-electron groundstate.

In this model, the only SP peak is identical to minus the ionization potential $\omega\QP = -I = E_{0}^{(2)} - E_{0}^{(1)}$, where $E_{0}^{(1)}$ is the one-electron ground-state energy. We omit all spin coordinates, as the exact GF is diagonal in spin for a non-magnetic groundstate. Summing over spins introduces a factor of 2, our results this far are not spin-summed.


Finally, we write SP contribution to the total two-electron ground-state energy,
\ben E_{0}\QP = \frac{1}{2}\left(\omega\QP f\QP + T\QP + \frac{\Delta v \Delta n\QP}{2}\right), \een
where $f\QP = \tr\{\rho\QP\}$ and $\Delta n\QP = n\QP_{2}-n\QP_{1}$. The MP contribution is defined analogously; together they sum to the total energy,
\ben E_{0}^{(2)} = \frac{1}{2}\left(\omega\QP  f\QP + \omega\SAT f\SAT  + T + \frac{\Delta v \Delta n}{2} \right),\een
where the full interacting kinetic energy is the sum
\ben T = T\QP + T\SAT.\een

\sec{One-shot $GW$ dimer self-energy}
To produce the $GW$ expressions we define the RPA polarization, of any $G$:
\ben P\RPA_{ \mathit{ij}\sigma}(\omega) = -i\int \frac{d\omega'}{2\pi}  G_{\mathit{ij}}(\omega')G_{ \mathit{ij}}(\omega+\omega'),  \een
which has a form analogous to the symmetric case. Considering a generic two-site tight-binding GF, as in App.~\ref{sec:TBGF},
\begin{multline} P\RPA_{\rm{s},\mathit{ij}\sigma}(\omega) =  \frac{n_{1}n_{2}}{4}\Big\{ \frac{(-1)^{i+j}}{\omega - 2t\s +i\delta} - \frac{(-1)^{i+j}}{\omega + 2t\s - i\delta} \Big\},\end{multline}
with screened interaction,
\ben W_{\rm{s},\mathit{ij}}(\omega) =  U\delta_{ij} + \frac{(-1)^{i+j}2U t\s n_{1}n_{2}}{(\omega -2t\eff + i\delta)(\omega + 2t\eff -i\delta)}, \een
where $t\eff = 2\sqrt{t\s^{2} + U t\s n_{1}n_{2} }$. Conveniently, the asymmetric screened interaction can be found from a simple rescaling of the inverse dielectric function $W_{ij}(\omega)/U$ computed by Romaniello and co-authors \cite{RGR09,RBR12} for the symmetric case : $t\to t\s$ and $U\to U n_{1} n_{2}$.

Finally we compute the one-shot $GW$ self-energy according to,
\ben \Sigma^{GW}_{ij}(\omega) = v_{{\rm{H}},ij} + i\int \frac{d\omega'}{2\pi}G_{{\rm s},ij}(\omega)W_{{\rm s},ij}(\omega+\omega'), \een
which yields,

\begin{multline} \Sigma^{GW}_{ij}(\omega) = \frac{U n_{i}\delta_{ij}}{2} + (-1)^{i+j}\frac{U^{2}t\s n_{1}n_{2}}{t\eff} \\ \times \Big\{\frac{\rho\s(n_{1},n_{2})_{ij}}{\omega - (\bar{v}\s - t_{1})-i\delta}  + \frac{\bar{\rho}\s(n_{1},n_{2})_{ij}}{\omega - (\bar{v}\s + t_{1})+i\delta}\Big\},  \label{eq:SigGW}\end{multline}
where $t_{1} = t\s + t\eff$ and all quantities depend on $n_{1,2}$ explicitly or through $\Delta v\s$. If the exact site-occupations and KS potential is used, then the self-energy is the one-shot Kohn-Sham self-energy. It can be understood as the one-shot self-energy from the best possible DFT calculation for the dimer. For the case of a one-shot Hartree-Fock calculation we take $\bar{v}\s \to U/2$, $\Delta n \to \Delta n\HF$, $\Delta v\s \to \Delta v\HF = \Delta v + U\Delta n\HF/2$, the corresponding HF parameters \cite{Carrascal_2015}.

\sec{One-shot $GW$ dimer GF}

The amplitudes of the one-shot $GW$ Green's function can be found directly from the Dyson equation,
\ben G^{GW} = \big[ G_{0}^{-1} - \Sigma^{GW} \big]^{-1}, \een
where $G_{0}$ is asymmetric tight-binding Green's function, which can be found by evaluating $G\s$ at the non-interacting site-occupations, i.e. $G_{0} = G\s(\Delta n = \Delta v /\bar{t}(\Delta v))$ and $\Sigma^{GW}$ is given by Eq.~(\ref{eq:SigGW}). We choose $\bar{v} = 0$ for convenience.

Formally we express the Green's function as a sum of matrices,
\begin{multline} G^{GW}(\omega) = \frac{X\QP}{\omega - \omega\QP - i\delta} + \frac{X\SAT}{\omega - \omega\SAT - i\delta} \\ + \frac{\bar{X}\QP}{\omega + \omega\QP + i\delta} + \frac{\bar{X}\SAT}{\omega + \omega\SAT + i\delta}, \end{multline}
where $\bar{X}\QS = \sigma_{y}X\QS\sigma_{y}$, and $X\QS$ has a cumbersome expression that can be computed in closed form. 
Here we only report $G^{GW}_{11}$ and $G^{GW}_{12}$ because entries of $X\QS$ can be constructed explicitly from only these elements. For the moment we ignore the time-ordering in the pole structure,
\ben G^{GW}_{11}(\omega) = \frac{(\omega - \frac{\Delta v\HF}{6})^{3} -a(\omega - \frac{\Delta v\HF}{6}) + b}{(\omega^{2}-(\omega\QP)^{2})(\omega^{2}-(\omega\SAT)^{2})},  \een
with
\ben \Delta v\HF = \Delta v + \frac{U \Delta n\HF}{2}, \een
where the HF density is the solution of $\Delta v\s(\Delta n) = \Delta v + U\Delta n/2$, the HF problem. The associated coefficients
\ben 
a = A + t_{1}^{2} + \frac{1}{12}(\Delta v\HF)^{2}, 
\een
\ben b = \frac{At_{1}\Delta n\HF}{2} + \Big(t_{1}^{2} - \frac{A}{2}\Big)\frac{\Delta v\HF}{3} -\frac{1}{4}\Big(\frac{\Delta v\HF}{3}\Big)^{3},
\een
are all functions of the HF density; either explicitly or implicitly through $\Delta v\HF$. Here $t_{1} = t\HF + t\eff$, with $t\HF = t\sqrt{1+(\Delta v\HF/2t)^{2}}$, $t\eff = 2\sqrt{(t\HF)^{2} + U t\HF n_{1}\HF n_{2}\HF}$, and $A = U^{2} n\HF_{1} n\HF_{2} t\HF /t\eff$.

The exact pole locations can be expressed in terms of these parameters,
\begin{multline} (\omega\QS)^{2} = A + \frac{1}{2}\left[t^{2} + t_{1}^{2}  + \Big(\frac{\Delta v\HF}{2}\Big)^{2}\right]^{2}  \\ \mp \Bigg[\Bigg(\frac{t^{2}-t_{1}^{2}}{2} + \frac{1}{2}\Big(\frac{\Delta v\HF}{2}\Big)^{2}\Bigg)^{2} \\ + A\Bigg((t^{2} + t_{1}^{2} )  + \Big(\frac{\Delta v\HF}{2}\Big)^{2} \\ -\frac{t_{1}}{2}\Big( 2t\sqrt{4-(\Delta n\HF)^{2}} + \Delta v\HF\Delta n \Big)\Bigg) \Bigg]^{1/2}, \end{multline}
and thus, the poles are functions of the HF density. Since the poles are clearly identifiable, the associated SP and MP spectral weights can determined via evaluation of the residues. For the SP matrix,
\ben X\QP_{11} = \frac{(\omega\QP - \frac{\Delta v\HF}{6})^{3} -a(\omega\QP - \frac{\Delta v\HF}{6}) + b}{2\omega\QP((\omega\QP)^{2}-(\omega\SAT)^{2})}, \een
and the $22$ element is found from simple substitution:
\ben X\QP_{22} = \frac{(\omega\QP + \frac{\Delta v\HF}{6})^{3} -a(\omega\QP + \frac{\Delta v\HF}{6}) - b}{2\omega\QP((\omega\QP)^{2}-(\omega\SAT)^{2})}, \een
which is equivalent to $\Delta v\HF \to - \Delta v\HF$. Conveniently, the many-particle counterparts are analogous:
\ben X\SAT_{11} = \frac{(\omega\SAT + \frac{\Delta v\HF}{6})^{3} -a(\omega\SAT + \frac{\Delta v\HF}{6}) - b}{2\omega\SAT((\omega\SAT)^{2}-(\omega\QP)^{2})}, \een
\ben X\SAT_{22} = \frac{(\omega\SAT - \frac{\Delta v\HF}{6})^{3} -a(\omega\SAT - \frac{\Delta v\HF}{6}) + b}{2\omega\SAT((\omega\SAT)^{2}-(\omega\QP)^{2})}. \een

To produce the off-diagonal elements of $X\QP_{12}$ we use
\ben G^{GW}_{12}(\omega) = \frac{-(t\omega^{2}-c^{2})}{(\omega^{2}-(\omega\QP)^{2})(\omega^{2}-(\omega\SAT)^{2})}, \een
where
\ben c = \sqrt{\frac{t_{1}}{2}\Big(2t_{1}t + A\sqrt{4-(\Delta n\HF)^{2}} \Big)},\een
and after computing the residue,
\ben X\QP_{12} = X\QP_{21} = \frac{-(t(\omega\QP)^{2} - c^{2})}{2\omega\QP((\omega\QP)^{2}-(\omega\SAT)^{2})}, \een
where the MP contribution is exactly $\omega\QP \to \omega\SAT$.

From $X\QS$ we define the approximate SP and MP occupations $\tilde{n}\QS_{i} = X\QS_{ii}$.  The KS expressions can be produced similarly, but they are slightly more complicated. Here we only report the one-shot GF beginning from a HF calculation. In the KS case, the constant $\bar{v}\s$ is designed to guarantee that the IP theorem is satisfied; this constant alters the locations of the poles and the spectral weights in a more complicated way.

\sec{Hubbard dimer one-shot and self-consistent $GW$ tables}

We report additional Hubbard dimer $\D\xc$ and associated spectral decompositions of the correlation energy. For the symmetric dimer ($\Delta v = 0$) each variant of $GW$ recovers the correct site occupations ($\Delta n = 0$). For two electrons with spin-singlet ground-state, this results in exactly reproducing the Hartree and exchange energies. This is because $U = -E\x/2$ for this case. As a result, the errors between the exact and approximate $\D\xc$ (and $E\xc$) are entirely due to correlation effects. In the asymmetric case, reported in the main text, the site-occupations are not correctly reproduced by the approximate methods, and thus they introduce errors to the Hartree and exchange energies.
\begin{table}[h]
  \centering
  \begin{tabular}{|c|c|c|c|c|c|c|c|c|c|c|c|c|c|c|c|}
    \hline
    \multicolumn{1}{|c}{} &
    \multicolumn{3}{c|}{Exact} &
    \multicolumn{3}{|c|}{one-shot} &
    \multicolumn{3}{|c|}{self-consistent} \\   
    \hline
    $U$  &  $\D\xc$ & $ \D\xc\QPI $ & $ \D\xc\STI$ &  $\D\xc$ & $ \D\xc\QPI $ & $ \D\xc\STI$ &  $\D\xc$ & $ \D\xc\QPI $ & $ \D\xc\STI$   \\
    \hline
   0.25 &  0 & 1.93 & -1.93 &  -1.14 & 2.54 & -3.68  & -15.4 & 2.58 & -18. \\
   \hline
   0.5 &  0 & 7.46 & -7.46 &   -6.01 & 7.25 & -13.3  & -27.4 & 7.65 & -35. \\
   \hline
   1 &  0 & 26.4 & -26.4 &   -25.4 & 17.3 & -42.7  & -46.9 & 19.8 & -66.7  \\
   \hline
   2 &  0 & 73.2 & -73.2 & -84.5 & 34.2 & -119. & -106. & 44.5 & -151. \\
   \hline
   4 & 0 & 138.\phantom{-0} & -138. & -225. & 56.4 & -282. & -310. & 89.2 & -400. \\
   \hline
  \end{tabular}
  
\vspace*{0.25 cm}

  \begin{tabular}{|c|c|c|c|c|c|c|c|c|c|c|c|c|c|c|c|}
    \hline
    \multicolumn{1}{|c}{} &
    \multicolumn{3}{c|}{Exact} &
    \multicolumn{3}{|c|}{one-shot} &
    \multicolumn{3}{|c|}{self-consistent} \\   
    \hline
    $U$  &  $E\c$ & $ E\c\QPI $ & $ E\c\STI$ &  $E\c$ & $ E\c\QPI $ & $ E\c\STI$ &  $E\c$ & $ E\c\QPI $ & $ E\c\STI$   \\
    \hline
   0.25 & -7.78 & -5.82 & -1.96 &  -12.6 & -8.81 & -3.74  & -26.2 &-8.17 &-18. \\
   \hline
   0.5 &  -30.8 & -22.9 & -7.92 &   -42.1 & -28.3 & -13.8  & -58.7 &-23.1 &-35.5 \\
   \hline
   1 &  -118. & -85.4 & -32.6 &    -127. & -81. & -46.3  & -121. & -51.8 &-69.7  \\
   \hline
   2 &  -414. & -280. & -134. &  -341. & -204. & -136. & -248. & -84.7 & -163. \\
   \hline
   4 & -1236. & -756. & -480 &  -807. & -460. & -347. & -512. & -76.6 &-436. \\
   \hline
   
  \end{tabular}
\caption{Decompositions for $\D\xc$ (top) and $E\c$ (bottom) for the exact and approximate symmetric dimer with $t=1/2$ and $\Delta v = 0$. For the self-consistent result we report the sum of the MP contributions. All energies in milliHartrees (mH).}
\label{tab:tab4}
\end{table}

In Table~\ref{tab:tab4} we report $G\xc$ and its decompositions with the IP contribution removed. We choose this partition because the magnitudes of the SP and MP contributions are reduced by a factor of 2. 

The symmetric dimer has $E\xc = \langle v\xc \rangle/2$ and so $\D\xc=0$, this implies,
\ben T\c = (\omega\QP-\omega\SAT)f\SAT, \een
where $\omega\QP-\omega\SAT = -2E_{0}(1)$ and $f\SAT$ is the MP occupations. One-shot and self-consistent $GW$ will never be able to produce this result, as they both predict an interaction dependent energy for the one-electron space ($\omega\QP-\omega\SAT$ changes with $U$), i.e., they have self-interaction error.

\sec{Implementation of self-consistent $GW$}
Self-consistent $GW$ results for the symmetric Hubbard dimer have been reported in Refs \cite{SLR21} and \cite{MZV22}. Fully self-consistent results on a frequency and time grid, without a pole purging scheme, are computationally non-trivial. Even for a two-site model, to capture the emerging "satellite series" requires many points to accurately resolve the details of the spectrum \cite{MZV22}. 

\begin{figure}[h]
\includegraphics[width=\linewidth]{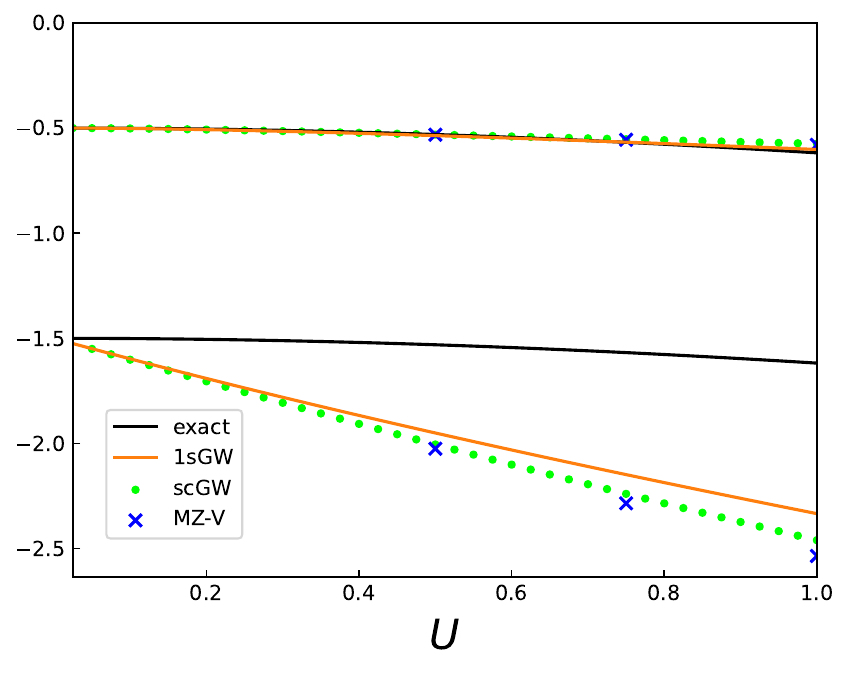}
\caption{Exact and approximate SP and MP energies as a function of interaction $U$. The upper lines are the SP energies while the lower are MP energies. We label the pole locations of \cite{MZV22} as "MZ-V" in blue.}
\label{fig:QPSATV}
\end{figure}

We instead report a finite pole version of self-consistent $GW$. At each iteration we retain an explicit two-pole form of the self-energy. This ensures a six-pole form of the $GW$ Green's function. Purging the poles allows us to exploit the meromorphic representations of all quantities, avoiding the need for frequency grids. For increasing $U$ values we find systematic deviations from the results of Mejuto-Zaera and Vl\v{c}ek, but we limit our analysis to small $U$ values, where $GW$ can be accurate and we find good agreement. In Fig. \ref{fig:QPSATV} we report the deviation of the single-particle and the first prominent many-particle pole position, we also include the relevant values from Ref\cite{MZV22}, and the one-shot results for the symmetric dimer. The deviation between our results with Mejuto-Zaera and Vl\v{c}ek are much smaller in magnitude than the errors between the one-shot result and our respective flavors of $GW$. Thus, for small and moderate $U$, our dimer results should be consistent with $GW$ results of Ref.~\cite{MZV22}. For large values of $U$, $GW$ is not expected to perform well, but we report the values from our finite pole scheme.

We have tested several pole truncation schemes and find that keeping a two-pole structure for $W$ and $\Sigma^{GW}$ produces converged total energies quickly for the full range of $U$ values computed. To aid convergence, for cases with moderate to strong correlation we generated the $n$th GF according to: $G^{(n)} = \gamma G^{(n-1)} + (1-\gamma) G^{(n-2)}$, with $\gamma = 0.67$. Our one-shot results are identical to Ref \cite{RGR09} in the symmetric limit $\Delta v \to 0$. 

\sec{$\G\xc$ for Coulombically interacting systems}

Huang and Umrigar report exact and KS energies for two-electron ions with nuclear charge $Z$ and the Hooke's atom system, which consists of two Coulombically interacting electrons confined in a harmonic oscillator with strength $\omega$ \cite{HU97}. Using these energies, we produce $\D\xc$ for a range of moderately and weakly correlated system parameters for Coulombically interacting electrons. 

For the case of a two-electron spin-singlet groundstate, the exchange energy of the KS system is exactly minus half the Hartree energy. The multiplicative exchange potential $v\x$ is a functional derivative of the exchange energy, and thus $v\x = -v\H/2$. While the full exchange self-energy evaluated with KS GF,
\ben \Sigma\x[G\s](\bx,\bx') = -\frac{\rho\s(\bx,\bx')}{\vert \br - \br'\vert}, \een
yields the KS exact exchange energy when it is traced with $G\s$,
\ben E\x = \frac{1}{2}\Tr\{\Sigma\x[G\s]G\s\} = -U/2, \een
and thus
\ben \D\x = \Tr\{(\Sigma\x[G\s] - v\x[n])G\s\} = 0. \een
Consequently, our tables contain only the correlation contribution $\D\c$. Table II of the main text and Table \ref{tab:Hooke} both show that $\D\xc$, $\D\c$ for this case, is generally smaller in magnitude than $E\xc$, its most analogous counter-part. We report the two-electron ion results in the main text. 
\begin{table}[h]
  \centering
  \begin{tabular}{|c|c|c|c|c|c|}
    \hline
    $\omega$ & $ E\xc $ & $ \langle v\xc\rangle/2 $ & $E\c$ & $T\c$ & $ \D\xc$ \\
    \hline
    0.1 & -0.2402 & -0.2311 & -0.02924 & 0.01643 & -0.009139 \\
    \hline
    1 & -0.7881 & -0.7757 & -0.04139 & 0.03393 & -0.0124 \\
    \hline
    4 & -1.587 & -1.574 & -0.04529 & 0.04097 & -0.01326 \\
    \hline
    10 & -2.515 & -2.502 & -0.04685 & 0.04397 & -0.01357 \\
    \hline
    100 & -7.972 & -7.958 & -0.04878 & 0.04781 & -0.01393 \\
    \hline
    400 & -15.95 & -15.94 & -0.04924 & 0.04875 & -0.01402 \\
    \hline
    1000 & -25.22 & -25.21 & -0.04941 & 0.0491 & -0.01404 \\
    \hline
  \end{tabular}
    \caption{Various X and C energies for the Hooke's atom system reported in Ref\cite{HU97}. The weakly correlated limit is approached as well strength $\omega \to \infty$. }
  \label{tab:Hooke}
\end{table}

Table \ref{tab:Hooke} shows the results for the Hooke's atom system. We find similar trends to the two-electron ion case, in the approach to the weakly-correlated limit ($\omega \to \infty$) $\D\xc$ converges to a finite value that is smaller in magnitude than $E\c \approx -T\c$ in this limit.

\bibliographystyle{apsrev4-2}
\bibliography{main}

\label{page:end}